\documentclass[aps,reprint,groupedaddress,superscriptaddress]{revtex4-1}
\usepackage{amsmath}
\usepackage{amssymb}
\usepackage{amsfonts}
\usepackage[dvipdfmx]{graphicx,color}
\usepackage{textgreek}
\usepackage{makecell,tabularx}
\setcellgapes{3pt}

\usepackage{amsfonts}
\usepackage[colorinlistoftodos]{todonotes}
\usepackage{multirow}

\usepackage[normalem]{ulem}
\usepackage{color}

\begin{document}

\title{Penetration of Rigid Rods, Flexible Rods, and Granular Jets \\ into Low-Density Granular Media}

\author{J.E. Ben\'itez-Zamudio}
\affiliation{Instituto de F\'isica, Benem\'erita Universidad Aut\'onoma de Puebla, A. P. J-48, Puebla 72570, Mexico \\
}

\author{S. Hidalgo-Caballero}
\affiliation{Gulliver, CNRS, ESPCI Paris, Universit\'e PSL, 75005 Paris, France, European Union \\
}

\author{F. Pacheco-V\'azquez}
\affiliation{Instituto de F\'isica, Benem\'erita Universidad Aut\'onoma de Puebla, A. P. J-48, Puebla 72570, Mexico \\
}

\date{\today}

\begin{abstract}
The penetration of projectiles into granular materials has been mainly studied using spherical intruders. Here we explore the dynamics of rods penetrating vertically in a two-dimensional granular bed composed of expanded polystyrene spheres.  The experiments were performed using rigid rods,  flexible rods and vertical arrays of non-cohesive particles, and the dynamics for the three cases was compared.  In contrast to the vertical penetration observed for a single spherical projectile, high speed videos reveal that  a rod rapidly deviates from its initial vertical direction due to inhomogeneities of the bed packing fraction. Then, the rod rotates due to the torque induced by the resistance force and follows a curved trajectory until be aligned horizontally at a final depth.  A short rod tends to deviate faster than a longer rod due to the smaller moment of inertia.  Moreover, long flexible rods always lose their vertical alignment and experience buckling, whereas rigid rods of the same size penetrate deeper before being deviated.  On the other hand,  experiments and molecular dynamics simulations show that a initially vertical array of grains also loses its verticality and stops adopting a final horizontal configuration.  The granular array penetrates considerably less than the rods of equivalent mass, and the stopping mechanism is based on vertical-to-horizontal momentum transfer during a collisional process of the constituting particles.
\end{abstract}


\maketitle

\section{Introduction}
\label{intro}
Pioneering studies focused on the resistance of a granular medium to projectile penetration were motivated by military applications \cite{Robins1742}.   Over the last decades, the study of granular drag against vertical penetration gained considerable attention \cite{ Uehara2003, Katsuragi2007, Seguin2009, Clark2013} due to its relevance in understanding cratering phenomena,  engineering applications,  spacecraft landing and locomotion \cite{Goldman2008,Pacheco2019,Maladen2009,Chen2013}.  Most penetration studies were focused on individual spherical impactors.  In contrast,   the vertical penetration of a group of intruders into a low-density granular bed was considered in Ref. \cite{Pacheco2010}.  In such scenario,  a collective dynamics was observed,  in which the disks lose the vertical alignment and repel each other due to the fluidization of the granular medium induced by the intruders,  until a final horizontal distribution of equilibrium is reached.  This first study in low-density granular materials motivated further research to explore terminal velocities of falling intruders \cite{Pacheco2011,Pacheco2017} and the effect of acceleration on the final depth simulating different gravities in extraterrestrial scenarios \cite{Altshuler2014}.  Other factors can deviate an intruder from its vertical penetration.  Recently,  it was shown that an imperfect cylinder with two halves of different roughnesses rotates toward the granular region adjacent to the smoother surface \cite{Espinoza2023}.  Subsequently,  the results were used to explain the penetration of fox and human skulls into light granular matter (simulating snow),  showing rotation in the direction of  the “smoother” half of the skull \cite{Altshuler2024}.  Thus,  material fluidization,  friction and intruder geometry play a fundamental role in describing the deviation from vertical penetration dynamics.  Of particular interest has been the study of drag acting on rods moving in a granular bed \cite{Arsalan2017,Furuta2019}. This is relevant in active intruders,  for instance, sand lizards use the motion of their elongated bodies to move more efficiently through sand producing a fluidization front \cite{Maladen2009}, which has been used to design bio-inspired lightweight robots \cite{Zhang2013}.  Thin rods, forced to penetrate into a granular medium,  buckle \cite{Seguin2018}. The critical depth for buckling depends strongly on the rod stiffness,  which is relevant for explaining the deformation of root growth of plants \cite{Kolb2012}.  However, the above studies were focused on rigid rods and shallow penetration in dense granular packings.

In this study,  we examine the penetration of rigid and flexible rods into a low-density granular medium.  In comparison to previous works focused on this geometry, the light granular material allows deep penetrations.  We show that the rods deviate from the vertical due to local variations of the granular packing.  Rigid rods are able to penetrate deeper than flexible rods, because the latter suffer buckling,  but in both cases,  the rods finally align horizontally due to the net torque induced by the resistance force.  We compare the results with experiments and molecular dynamics simulations focused on a falling granular jet.  Both, the 2D experiments and simulations show that the final shape of the impacting jet adopts a quasi-horizontal distribution, as a consequence of a continuous change in the local density of the granular medium, which modifies the effective drag forces acting on the impactors \cite{Pacheco2010}. 

\section{Experimental setup}

The 2D experiments were performed using a vertical Hele-Shaw cell of 1 m$^2$ transparent glass walls separated 5.0 mm by a frame made on acrylic.  The cell was filled from the top by pouring expanded polystyrene spheres of diameter $d=4.7\pm 0.1$ mm,  and density $\rho_g = 0.031\pm 0.005$ g/cm$^3$, forming a monolayer.  The granular jets consisted of $N$ steel beads of 4.7 mm in diameter,  4.4 g of mass, and density $\rho_s=7.5$ g/cm$^3$.   To built the rods,  we used the same kind of steel beads and $N$ of them were glued with epoxy putty to produce rigid bars. The flexible rods were built using $N$ neodymium spherical magnets of 4.7 mm diameter  and 4.8 g of mass.  Both rigid and flexible rods of $N$ spheres were ensured to have the same mass using the appropriate amount of epoxy putty.   A rail was mounted  on the top of the cell to release the intruders vertically aligned  from a given height $H_0$ (see Fig. 1a).  After each experiment,  the intruder can be easily extracted from the cell using a magnet, and the cell (mounted on a frame that allows its vertical rotation) is inverted to displace the granular material and erase the history of the previous penetration.  The upper surface of the granular bed was kept always at the same horizontal level and the packing fraction obtained following this procedure was estimated to be $\phi \sim 0.66\pm0.02$.  
High-speed videos were recorded at 1000 fps and analyzed using the software ImageJ and Tracker (with a spatial resolution of 1280 x 1024 pixels,  $\sim 1$  pix/mm).  The system was illuminated from behind to make a clear contrast between the metallic intruders and the partially translucent polystyrene particles, which facilitate particle tracking and local packing density estimations.  The trajectories were visualized by subtracting consecutive frames of the videos, which also allows  the visualization of the fluidized regions of granular material.

\begin{figure}
\includegraphics[width=0.47\textwidth]{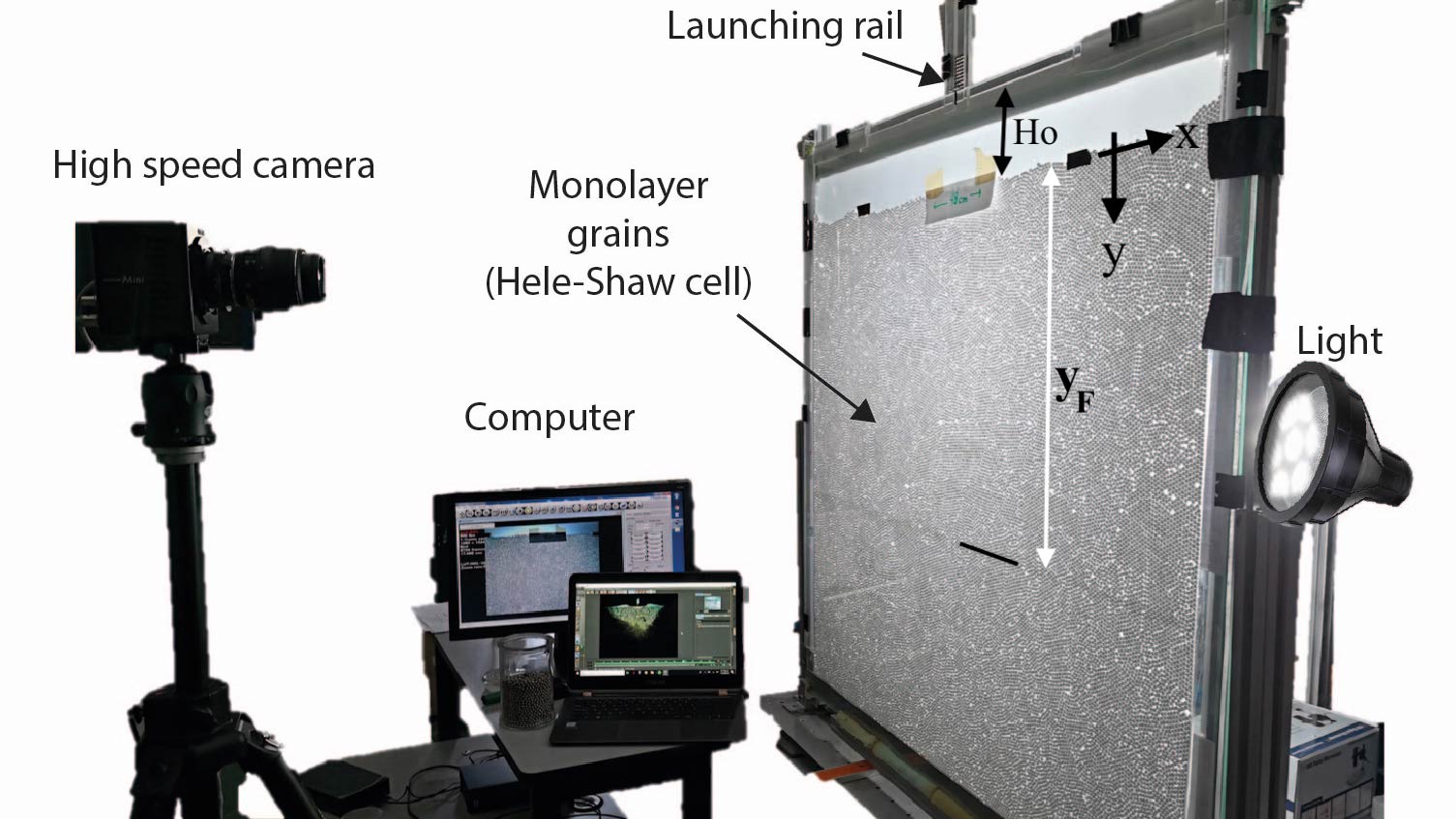}
\centering
\caption{Experimental setup. A Hele-Shaw cell of 1m$^2$ and 5 mm thick was filled with a monolayer of expanded polystyrene spheres. Then, different arrays were released from the top and the dynamics was filmed with a high-speed camera at 1000 fps.}
\end{figure}

\section{Results and discussion.}

\begin{figure}[h]
\centering
\includegraphics[width=1\columnwidth,clip]{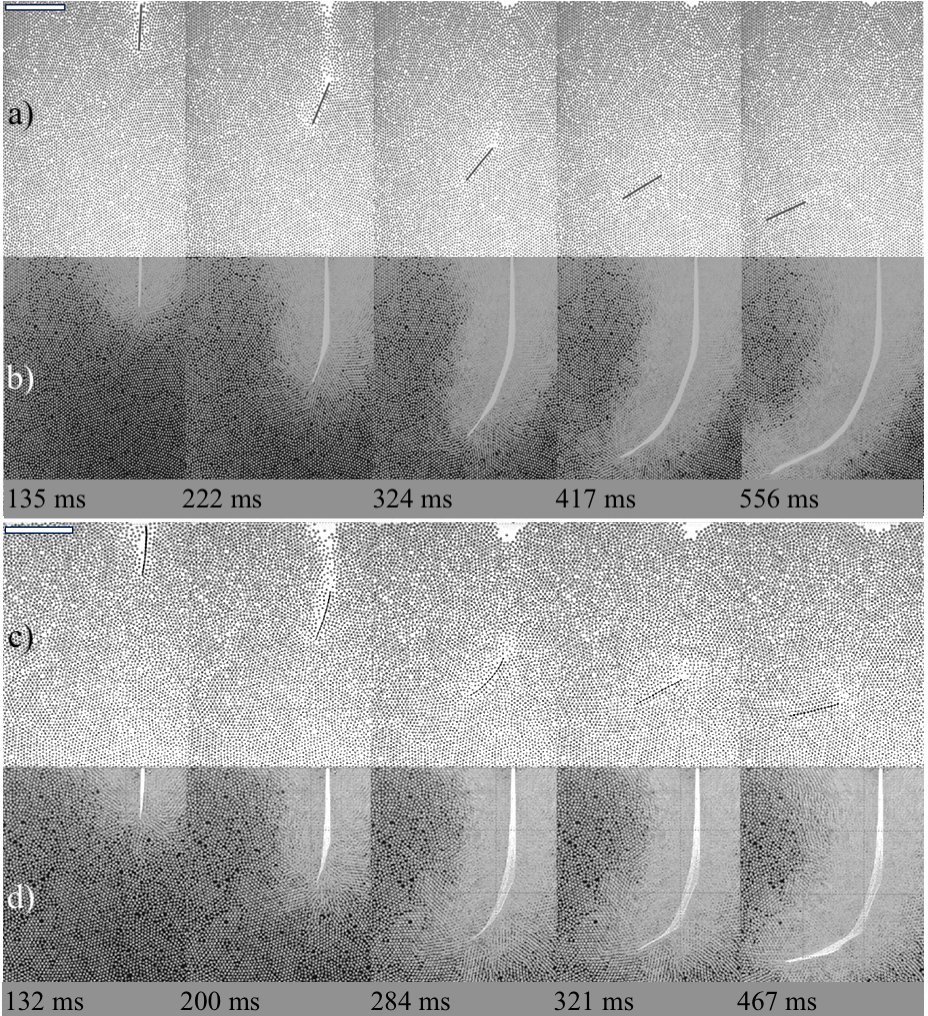}
\caption{Comparison of the penetration of a-b) rigid and c-d) flexible rods ($l= 7. 1$ cm,  $m=7.2 $ g) falling vertically into the 2D granular bed.  The intruder path and material fluidization is visualized.  Scale bar=10cm.}
\label{fig-2}       
\end{figure}

Figures \ref{fig-2}a-c show snapshots of the penetration of rigid and flexible rods of equal length  and mass composed of 15 beads.  First, the intruder penetrates vertically and fluidizes locally the granular bed.  In both cases, the rods start to lose the vertical alignment and tend to rotate towards a horizontal configuration.  For comparison reasons,  the rotation towards the left is shown in the snapshots for both cases,  but  the deviation can occur randomly towards the left or right and,  statistically,  there was no preferential side.  If the rod falls slightly tilted towards one side, the intruder will rotate in the same sense during its penetration until reaching the horizontal alignment.  Figures \ref{fig-2}b-d show the corresponding montages after consecutive image subtraction that allow to visualize the path of each intruder and the fluidized regions of the granular bed.   For short rods,  the penetration is very similar in both cases because the stiffness of the flexible rod is enough to keep it almost straight at shallow depths. 

\begin{figure}[h]
\centering
\includegraphics[width=1\columnwidth,clip]{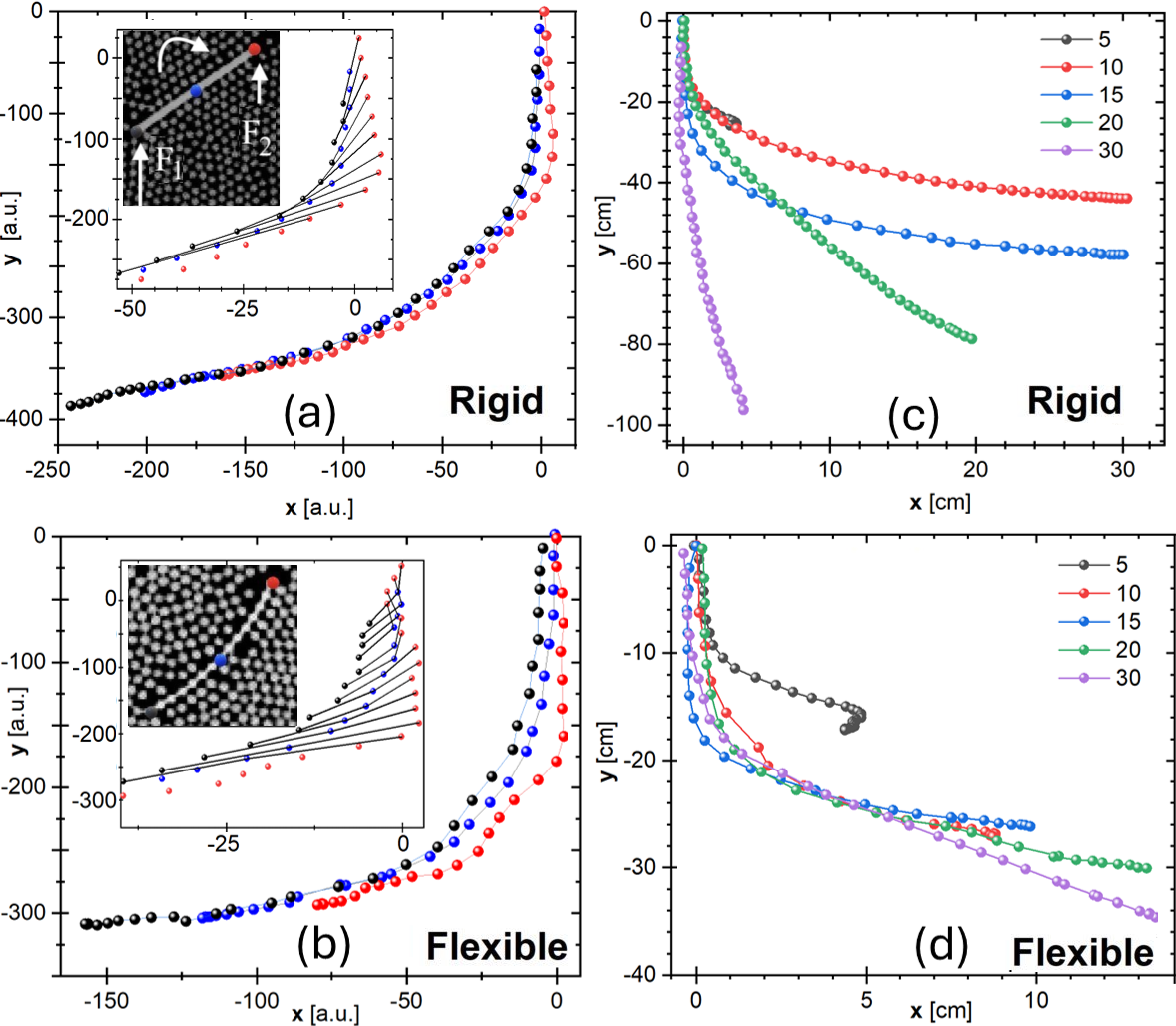}
\caption{Individual paths of a-c) rigid and b-d) flexible rods depending on the length $l$ (N spheres).  Insets show enlarged (rescaled) views of the paths for the three indicated coloured dots used to track the rotation of the rods.  (x,y) represent coordinates. }
\label{fig-3}       
\end{figure}

Figure \ref{fig-3} shows the (x,y) paths of the intruders for different rod lengths (number of particles $N$). To obtain the paths, the positions of the tip, center and tail of the rod were tracked (see color dots in the snapshots shown in the insets in Figs.  \ref{fig-3}a,b).  The three points fall in a straight line for a rigid rod;  in contrast,  the points are not necessarily aligned for the flexible rod that allows buckling, with the tail (red path) always bending below the center (blue) and tip (gray) paths.,  For the rigid case,  shorter rods are easier to deviate.  As the length of the rigid rod increases, the intruder goes deeper in the bed, and the deviation from the vertical is less notorious (Fig.  \ref{fig-3}c).  This can be explained considering that a larger rod has larger moment of inertia and is more difficult to deviate by asymmetries in the force chains network of the granular packing,  which become stronger at greater depths and eventually divert the intruder.  However,  flexible rods  buckle and deviate from the vertical,  penetrating considerably less than the rigid rod,  finishing always nearly horizontally aligned (Fig. \ref{fig-3}d). In fact,  longer flexible rods are more sensitive to local fluctuations of the packing fraction and the buckling occurs at shallow depths.  To explain the rotation, let us focus on the inset of Fig. \ref{fig-3}a.  The tip of the rod experiences a greater resistance force $F_1$ by the static grains than the force $F_2$ acting on the tail moving in the region fluidized by the tip. Thus, the net torque with respect to the center of mass causes the intruder to rotate,  as indicated by the curved arrow.  As the intruder aligns horizontally,  the torques equilibrate and the rod stop rotating.

\begin{figure}[h]
\centering
\includegraphics[width=1\columnwidth,clip]{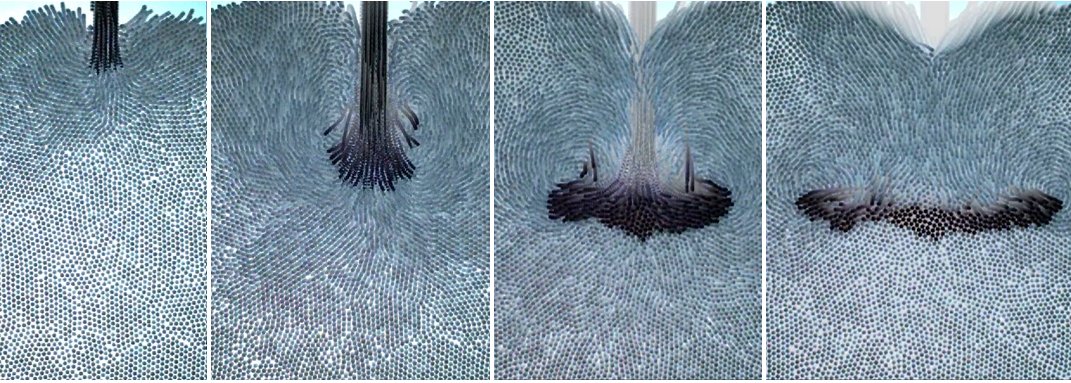}
\caption{Granular jet  of $25\times5$ steel spheres falling vertically into the granular bed stop in horizontal configuration. }
\label{fig-4}       
\end{figure}

For comparison,  we performed additional experiments with granular jets,  a column composed of $N$ non-cohesive steel beads falling under gravity.  Snapshots in Fig.  \ref{fig-4} show a vertical jet (N=125) entering the granular medium. The beads stop at a given depth and are dispersed laterally by collisions, forming also a horizontal arrangement. 

\begin{figure}[h]
\centering
\includegraphics[width=0.9\columnwidth,clip]{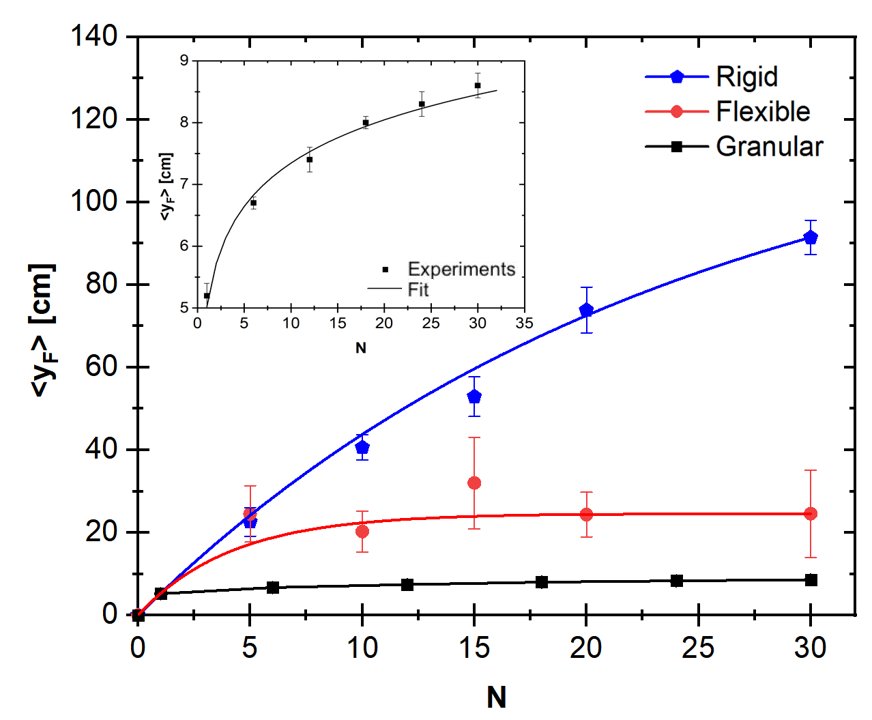}
\caption{$y_F$ vs $N$ for rigid rods, flexible rods and granular jets of the same length.   For clarity, $y_F$ vs $N$ is also shown in the inset for the granular jet.  Solid symbols represent the mean value and the standard deviation of five measurements,  and lines  the fits of the form: $y_F=A(1-e^{-N/N^*})$. }
\label{fig-5}       
\end{figure}

Figure \ref{fig-5} shows the final penetration depth $y_F$ as a function of the intruders length (in terms of $N$) for the three kinds of intruders falling from $H_0= 15$ cm.  As discussed above,  short rigid and short flexible rods ($N \leq 5$) behave similarly and reach the same depth,  but larger rigid rods penetrate deeper.  Note also that rods sink deeper than granular jets of the same size.  It can be observed in the inset that $y_F\sim 5$ cm for a single steel bead,  and this depth increases with the number of beads constituting the jet,  saturating at $y_{\infty} \sim 10$ cm, half of the saturation value observed for flexible rods.  A tendency to saturation is also observed for rigid rods (but deeper penetrations cannot be analyzed in our 1 m$^2$ cell).  Accordingly,  a good fit of the experimental data was obtained with an equation of the form $y_F=y_{\infty}(1-e^{-N/N*})$, see solid lines in Figure \ref{fig-5}. 

\section{Numerical Simulations}
\label{sec2}

\begin{figure}
\includegraphics[width=\columnwidth,clip]{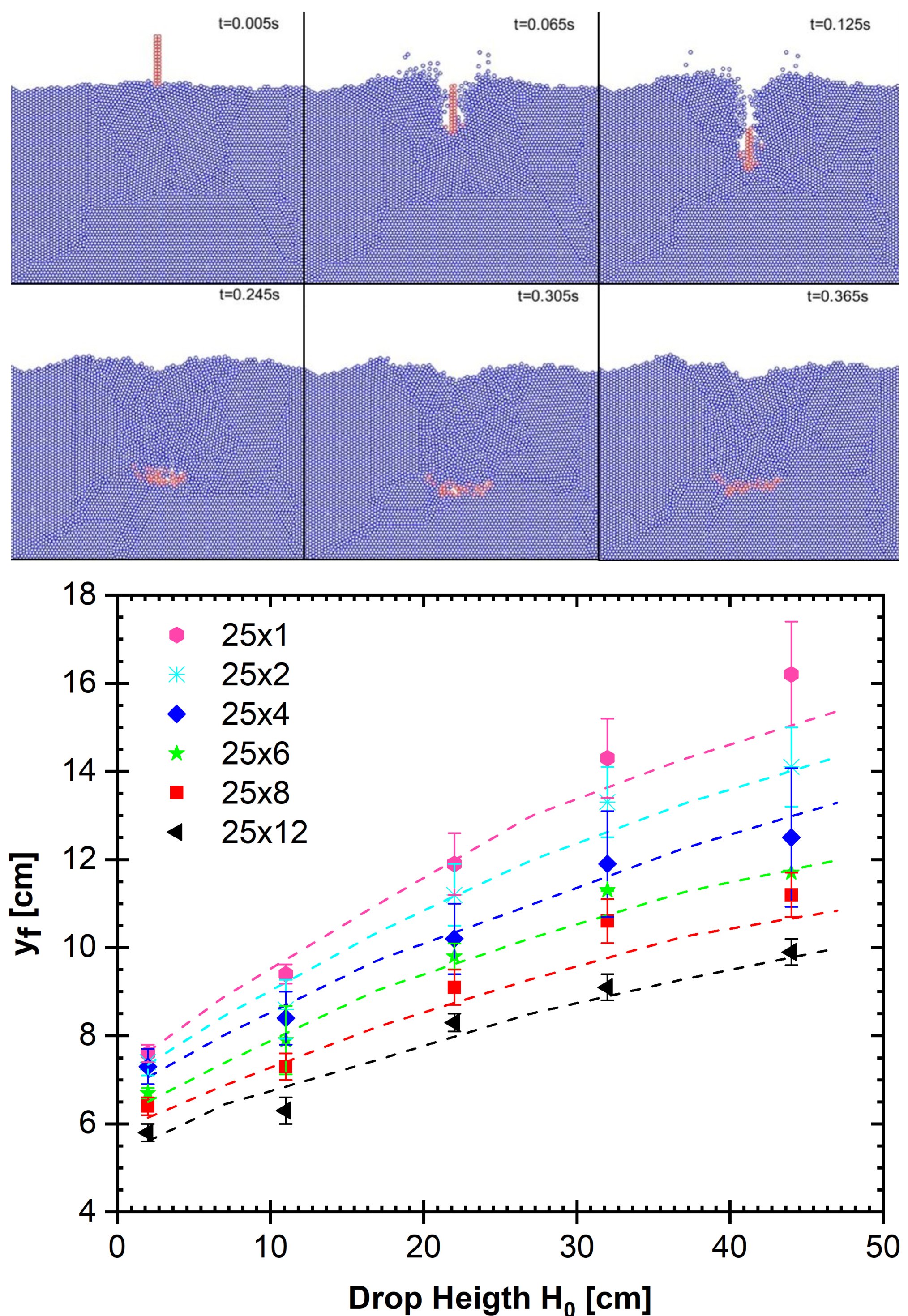}
\centering
\caption{Snapshots of simulations showing the entrance of an array of 15$\times$2 beads into a bed composed of 6000 expanded polystyrene particles  ($H_0=30$ cm.)  The plot shows the final depth $y_F$ reached by granular jets formed by a$\times$b spheres, where a=25 spheres are aligned vertically in b columns.  Solid points represent experiments and dashed lines numerical results for different drop heights $H_0$. }
\label{fig6}
\end{figure}

Finally,  molecular dynamics simulations were implemented in MatLab$^\circledR$ using the Velocity-Verlet algorithm for the case of granular jets.  This scheme is obtained from Taylor expansions for the positions and velocities\cite{frenkel2001}, \cite{bell2005}. At the time $t+\Delta t$ both are expressed respectively as: 
${\bf r} (t + \Delta t) = {\bf r} (t) + {\bf v} (t) \Delta t +
  {\bf a} (t) \Delta t^2 /2 $ and
${\bf v} (t + \Delta t) = {\bf v} (t) + \left( {\bf a} (t)   +   {\bf a} (t + \Delta t)\right) \Delta t /2$,
where  $a(t)$  is  the total force per mass acting on the particle. Here, we considered the particle weight, the normal force and the friction from the glass-particle interaction. According to the Hertz-law,  one term of the normal force $f_n$ depends on the particle Young's modulus $E$, its  radius $R$ and the particle deformation $\delta$ as  $f_{n_1}\sim E \sqrt{R} \delta^{3/2}$. The values of the Young's modulus were taken from \cite{youngs} and for steel and expanded polystyrene  these are $E_1=200$ GPa and $E_2=5$ MPa, respectively.  A second term of $f_n$ is related to the normal dissipation of energy and it depends on the deformation rate, thus, on the relative velocities of two colliding particles i, j: $\bf{v_r}={v_i}-{v_j}$, then $f_{n_2}= \alpha v_r$. Particle rotation was neglected and a damping force was introduced to model the particle-wall interaction,  $f_w=\beta  v$. The constants $\alpha$ and $\beta$ represent the dissipation terms among  the granular bed and due to wall contact, respectively. These parameters were estimated from the experimental stopping time and the final depth of a single impactor. By using the above  equations,  with $\Delta t= 5\mu s$, we computed the positions and velocities at each step of time, allowing us to follow the evolution of the system. Figure \ref{fig6} exemplifies the snapshots obtained numerically for the impact of a jet of 15$\times$2 particles (in red color) released at 30 cm above the granular bed (blue particles).  As in the experiments,  the granular jet fluidizes the bed and finishes in a horizontal configuration.  The effect of number of columns, size of the column and drop height are preliminarly shown in Fig. \ref{fig6}, comparing well experiments with simulations,  but this data will be explored more deeply in a future work.  MD simulations will be also used to explore the dynamics of rigid and flexible rods.  Since the effect of confinement can be crucial in all the observed dynamics,  the equivalent 3D system will be studied using X-ray tomography.

\section{Conclusions}
\label{sec4}

We have explored experimentally the vertical penetration of rods and granular jets into a low-density granular material.  In all cases,  the intruders deviate from the vertical and penetrate into the bed until reaching a final horizontal configuration.  For rigid rods,  the deviation begins when the rod is slightly tilted towards one side due to inhomogeneities that characterize a granular packing.  Once the tilting occurs, the rotation of the rod is amplified by drag-induced torques until reaching the horizontal alignment.  For flexible rods,  buckling also affects the penetration dynamics and the rods stop horizontally at shallow depths.  For granular jets, the first falling particles are stopped by the granular medium,  and upper particles collide with the former,  being scattered laterally.  This process results in a final horizontal distribution.  

\section{Acknowledgments}
\label{sec5}
This research was supported by CONAHCyT Mexico (now SECIHTI) and VIEP-BUAP Project 2025.


\begin{thebibliography}{}
\bibitem{Robins1742} B. Robins,  New Principles of Gunnery (Wright, London, 1742)

\bibitem{Uehara2003}
J. S.  Uehara,  M.A. Ambroso, R.P.  Ojha,  \& D.J. Durian, . 
Phys.  Rev.  Lett {\bf 90}(19), 194301 (2003).

\bibitem{Katsuragi2007}
H. Katsuragi,  D.J. Durian, 
Nat. Phys.  {\bf 3}(6), 420 (2007)

\bibitem{Seguin2009}
A. Seguin,  Y. Bertho,  P.  Gondret  \& J. Crassous,  
EPL {\bf 88}(4), 44002 (2009).

\bibitem{Clark2013}
A. H. Clark,  R.P. Behringer, 
EPL {\bf 101}, 64001 (2013).

\bibitem{Pacheco2019} F.  Pacheco-Vázquez, 
Phys. Rev. Lett.  {\bf 122}, 164501 (2019).

\bibitem{Chen2013}Chen Li et.  al.  
Science {\bf 339},1408-1412 (2013). 

\bibitem{Maladen2009}R.  D. Maladen et al. , 
Science {\bf 325}, 314-318 (2009). 


\bibitem{Goldman2008}
D. I. Goldman,  P.  Umbanhowar, 
Phys.  Rev.  E,  {\bf 77}(2), 021308 (2008).

\bibitem{Pacheco2010}
F. Pacheco-Vázquez and J. C.  Ruiz-Suárez, 
Nat Commun {\bf 1}, 123 (2010).

\bibitem{Pacheco2011}
F. Pacheco-Vázquez,  \textit{et. al .} 
Phys. Rev. Lett.  {\bf 106}(21), 218001 (2011).


\bibitem{Pacheco2017}
L.A. López-Rodríguez \& F. Pacheco-Vázquez,  
Phys. Rev.  E, 96(3), 030901 (2017).

\bibitem{Altshuler2014}
E. Altshuler,  \textit{et. al .}  
Geophys. Res. Lett.  {\bf 41}, 3032-3037 (2014).

\bibitem{Espinoza2023} M. Espinosa et al. ,  
Sci. Adv. {\bf 9},eadf6243(2023).

\bibitem{Altshuler2024} L. Mart\'inez-Ort\'iz, R. Pupo-Santos ,P.  Altshuler, and E. Altshuler,  
PNAS  {\bf 121} (33) (2024). 

\bibitem{Arsalan2017} M.  Arsalan \textit{et. al.},  
Advances in Service and Industrial Robotics.  RAAD 2017. Mechanisms and Machine Science 49. Springer, Cham.

\bibitem{Furuta2019} T. Furuta,  S. Kumar,  K.A.  Reddy,  H. Niiya,  H.  Katsuragi
New J.  Phys {\bf 21} (2), 023001 (2019).

\bibitem{Zhang2013} T. Zhang,  et. al. 
Int.  J.  Robotics Research {\bf 32}(7), 859-869 (2013).

\bibitem{Seguin2018} A. Seguin,P. Grondet,  
PhysRevE {\bf 98}, 012906 (2018).

\bibitem{Kolb2012} E. Kolb, C.  Hartmann \& P. Genet,  
P. Plant Soil {\bf 360}, 19-35 (2012).

\bibitem{frenkel2001}
D. Frenkel and B. Smit,  Understanding molecular simulation: from algorithms to applications (1996, Academic Press).

\bibitem{bell2005}
N. Bell,  Y. Yu, and P.J.  Mucha (2005).  In Proceedings of the 2005 ACM SIGGRAPH/Eurographics symposium on Computer animation (pp. 77-86). ACM.

\bibitem{youngs}
"Elastic Properties and Young Modulus for some Materials". The Engineering ToolBox.  (2012-01-06).\\


\end{thebibliography}
\end{document}